\documentclass[12pt]{article}
\usepackage{graphicx}
\usepackage{amssymb,amsmath,amsfonts,palatino,amsthm}
\usepackage{amssymb}
\usepackage{epstopdf}
\newcommand{\beq}{\begin{equation}}
\newcommand{\eeq}{\end{equation}}

\newcommand{\f}{\begin{equation}}
\newcommand{\ff}{\end{equation}}

\newcommand\be{\begin{equation}}
\newcommand\ba{\begin{eqnarray}}
\newcommand\ee{\end{equation}}
\newcommand\ea{\end{eqnarray}}

\begin{document}

\title{ Enhanced color gauge invariance and a new di-photon state at the LHC \\}
\author{Stephon Alexander\thanks{stephon.alexander@dartmouth.edu} , 
\\
Department of Physics \\
Brown University, Providence, R.I.  USA
\\
Lee Smolin\thanks{lsmolin@perimeterinstitute.ca} 
\\
Perimeter Institute for Theoretical Physics,\\
31 Caroline Street North, Waterloo, Ontario N2J 2Y5, Canada}
\maketitle

\begin{abstract}
\noindent 
We propose to interpret  the possible resonance seen in di-photons at the $LHC$ at $750 \ GeV$ as a bound 
state of a new pair of heavy gluons associated with an enhanced color gauge invariance.  
These have a conservation law which enforces their production and decay in pairs and hence requires that the leading coupling to 
quarks is quadratically through a dimension $5$ operator.  

One way to realize these hypotheses is if the $SU(3)$ color gauge invariance is enhanced to $SL(3,C)$, while at the same time promoting the internal metric, which picks out what is a unitary transformation, to a dynamical degree of freedom.   This theory was 
first proposed by Cahill\cite{Cahill}, Dell\cite{Dell}, Kim and Zee\cite{Zee}, and Julia and Luciani\cite{Julia}.  The dynamical internal metric spontaneously 
breaks $SL(3,C)$ to $SU(3)$ giving a mass to the vector bosons associated with the generators in $SL(3,C)/SU(3)$. The coupling to the internal metric also ensures that the energy is bounded from below.   
The new state is produced by gluon fusion and decays to  a pair of photons via a coupling to a new set of vector quarks allowed by the symmetries.

\end{abstract}


\newpage

\section{Introduction}

Recently the ATLAS and CMS collaborations have found an excess of diphoton events with a $3.9\sigma$ and $2.6\sigma$ stastistical significance respectively\cite{ATLAS,CMS}.  The data suggests a 750 GeV resonance in the diphoton invariant mass spectrum.  
This excess may turn out to be a statistical fluke, but it is possible that it is the result of physics beyond the standard model, in particular, a new state.  

In this note we suggest that the new state could arise as the bound state of a pair of heavy gluons (i.e. a color octet of massive vector mesons) if their dynamics is subject to a conservation law that requires they be created and annihilated in pairs.

There is a natural framework for this hypothesis.  A  vector octet with the required properties can arise from an enhancement of the $SU(3)$ color gauge symmetry to an $SL(3,C)$ gauge symmetry\cite{Cahill,Dell,Zee,Julia}.   This provides a highly constrained model which we will suggest may be consistent with what we know of the new state.

Thus far, a variety of models have been put forward to explain what the new state could be\footnote{For a small, unsystematic sample, see \cite{diagonal,MattS1,random}.}.  These have a range of motivations.  But note that the one thing that we can be sure must happen beyond the standard model is a unification with gravity.  It is then interesting that the extension of the color gauge group we describe is intimately tied with an idea as to how the standard model can be united with gravity\cite{Dell}.     

Indeed, the motivation for the extended gauge symmetry comes from an analogy to the equivalence principle.  
In $SU(N)$ Yang-Mills theory, the degree of freedom is a connection.  But note that the theory implicitly depends on the choice of a  metric on the internal space, $q$, which defines what a unitary transformation is.  In Yang-Mills theory this internal metric is fixed and non-dynamical.  Meanwhile, gravity is usually formulated as a dynamical theory of a spacetime metric, $g_{\mu \nu}$,  whose connection 
$\Gamma_{\mu \nu }^\lambda$ is a slave to it.  

The key is to understand that both metrics express a symmetry breaking\cite{Dell}.  In general relativity, the metric distinguishes the local Lorentz frames, carried by freely falling observers.  This  breaks local $GL(4,R)$ invariance of the tangent space to $SO(3,1)$, which are the local Lorentz transformations that preserve $g_{\mu \nu}$.  In Yang Mills theory the internal metric, $q$, distinguishes the local $SU(N)$ invariance from a more general $SL(N,C)$ invariance.  

In general relativity, the equivalence principle asserts that you  can always go to a freely falling frame by picking a gauge that reduces the metric locally to the Lorentz metric.  By analogy, if we wish to extend the equivalence principle to Yang-Mills theory, we can extend the gauge invariance of that theory from $SU(N)$ to $SL(N,C)$, and simultaneously promote the internal metric $q$ to a dynamical degree of freedom.   As in general relativity, we can always fix the gauge such that $q=1$. 

The  extension of Yang-Mills theory in which the gauge group is enhanced to $SL(N,C)$, while the internal metric $q$ becomes dynamical, was invented independently by Cahill\cite{Cahill},  Dell\cite{Dell}, Kim and Zee\cite{Zee}, and Julia and Luciani\cite{Julia}.  The quantum theory was studied in detail in \cite{JJL,ren,others-q}.

There are three main results which are possibly relevant for particle physics.  First, there is a natural, indeed mandatory, mechanism for spontaneous symmetry breaking, by which $SL(N,C)$ gauge symmetry breaks spontaneously to $SU(N)$.  The gauge fields in the non-compact directions $\frac{SL(N,C)}{SU(N)}$ get masses, $M$. We call these the $E$-bosons, for enhanced gauge symmetry.   There is no higgs field, rather the longitudinal modes of the gauge fields valued in $\frac{SL(N,C)}{SU(N)}$``eat" the degrees of freedom of the internal metric.  The mass scale $M$ arises because, as in general relativity, the metric is dimensionless and its dynamics at lowest order is governed by a dimension $2$ term.   Below that mass scale we have  $SU(N)$ Yang-Mills theory.  

Second, the Hamiltonian of the theory is bounded from below, because of the role the dynamical internal 
metric plays in the action\cite{JJL}.   Had we just extended $SU(N)$ to $SL(N,C)$ in the Yang-Mills action, this would not be the case and the theory would be unstable,  because the Killing form of non-compact groups is not positive 
definite\footnote{As Dell suggested in \cite{Dell}, this extension of the equivalence principle may be the key to unifying gravity and Yang-Mills theory.
One does this by embedding the Einstein-Yang-Mills theory in a new theory where both metrics and both connections are dynamical\cite{Dell}.  This idea that the metric and connection are independent has turned out to be key for quantum gravity\cite{Plebanski,Ashtekar} and plays a role in some supergravity models\cite{Julia}, but here we will be concerned only with its implications for elementary particle physics. }.  

Third, the theory incorporates a new parity symmetry, which is called $E$-parity, which implies that the massive vector mesons  can only be created or annihilated in pairs.

To summarize, the resulting theory describes an $SU(N)$ Yang-Mills theory  coupled consistently to a  set of massive vector bosons in the adjoint representation, which represent the $SL(N,C)/SU(N)$  generators.  This expresses a new realization of dynamical gauge symmetry breaking in which the mass of a vector meson arises when the longitudinal mode of the gauge fields associated to  the $SL(N,C)/SU(N)$ generators absorbs the degrees of freedom of the internal metric\footnote{A possible application to electroweak symmetry breaking was explored in \cite{DellSmolin}.}.    
The theory has overall local $SL(N,C)$ gauge symmetry and global $E$-parity, so the interactions are tightly constrained. 
At the level of effective field theory relevant for the $LHC$ observations there are two new free parameters:  the mass $M$ and a new gauge coupling constant $f^2$.

Our proposal is that the new state possibly seen at the $LHC$ is a bound state consisting of a pair of these new vector bosons, which arise from enhancing color $SU(3)$ to $SL(3,C)$,  confined into a color and Lorentz singlet.  This determines the mass $M \approx 375 \  GeV$.  

Because of $E$ parity, this bound state does not couple easily to photons, but the rate to photons can be enhanced to agree with the possible new observations if one couples it to new heavy vector quarks, as shown in Figure 1.

This has immediate implications for a few further states which must exist if this interpretation is correct.  One is a single $E$ coloured singlet, which can be thought of as a bound state of an $E$ meson with a gluon, of roughly mass $M$.  This is $E$ parity odd and 
while subject to annihilation in pairs is otherwise  long lived.  This is a possible candidate for a component of strongly interacting, dark matter, and would have to evade limits on such.   (Thus it is likely at best only a small component of the dark matter.) There must also be colour singlets of a $E$ boson bound with a quark-antiquark pair that decay like very heavy mesons.  Additionally,  we expect a spectrum of excited states of the bound pair, with spacings analogous to those of quarkonium, of order $\alpha_S^2 M$.  
These will decay to the ground state singlet through emission of gluons.  

In addition, if the same enhancement occurs for the electroweak $SU(2) \times U(1)$ we expect heavy photons, $W$'s and $Z$''s.   

At this early stage, any proposal for the interpretation of the new resonance is likely to be wrong, and we must caution that, given the look-elsewhere effect, even our confidence that this state is real cannot be high.  But given the motivation in terms of an extension of the equivalence principle, this is a possibility to be explored.   

We should note right away that the theory as originally written down 
in \cite{Cahill,Dell,Zee,Julia} is not perturbatively renormalizable\cite{JJL}.    
It is best regarded as an effective field theory, and this is how we will treat it here, postponing the issue of its ultraviolet completion.  

One completion which has been studied comes by writing the full action containing terms of dimension four or less\cite{ren}.  This action is perturbatively renormalizable and ghost free, but there are so far unresolved issues with the stability of the perturbative vacuum\cite{ren}. 
These issues may be resolvable as they do not, as in gravity, involve higher derivative instabilities, but we will not attempt to resolve them here\footnote{Possible mechanisms to resolve them include asymptotic safety\cite{AS} and the Lee-Wick mechanism\cite{LW}. Or it might require completion in some very different theory.}. The difficulties with stability can however be pushed off to an arbitrary higher mass scale, $\sqrt{h}M$ where $h >> 1$ is a free parameter of the fully renormalizable action. Thus we can treat the theory relevant for the physics at and above  scale $M$, as we do the Einstein theory, as an effective low energy action, in which case we deal with a stable  theory whose Hamiltonian is positive definite\cite{JJL}.     

We should also note that our theory has some similarities with-but also key differences from-the proposal\cite{diagonal} that the $QCD$ color gauge symmetry is the diagonal subgroup of a local $SU(3) \oplus SU(3)$ gauge symmetry, which is spontaneously broken to $SU(3)$.  There the heavy gluons live in the coset $\frac{SU(3) \oplus SU(3)}{SU(3)_d}$ in contrast to the theory described here, where they live in 
$\frac{SL(3,C)}{SU(3)}$.  As a result, they have no conservation law analogous to $E$ parity.

In the next section we write down the effective action governing this enhancement.   Section 3 reviews the basic physics of these models as preparation for section 4 when we  discuss briefly the possible implications for recent experimental results.

\section{The extended gauge theory model}

We extend the colour $SU(3)$ gauge theory to an $SL(3,C)$ gauge theory, following Cahill\cite{Cahill},  Dell\cite{Dell} and others\cite{Zee,Julia,JJL,ren}. 
A spontaneous breaking of $SL(3,C)$ to
$SU(3)$ is implemented by making the hermitian metric on the fundamental three dimensional representation dynamical.  This metric will be denoted $q_{\alpha \beta}$.  The $SL(3,C)$ connection is $\omega_{\mu \alpha}^{\ \beta}$ and its curvature or field strength is $F_{\mu \nu \alpha}^{\ \beta}$.   The theory is described by an effective action
\f
S= \int d^4 x {\cal L}
\ff
where, 
\begin{eqnarray}
{\cal L}^{eff}  &=&  -\frac{1}{4g^2_{1}} Tr   [q^{-1} F_{\mu \nu}^\dagger q F^{\mu \nu} ]
-\frac{1}{8g^{ 2}_{2}} Tr  [ F_{\mu \nu}  F^{\mu \nu} + F_{\mu \nu}^\dagger  F^{\dagger \mu \nu}  ]
\nonumber \\
&&+ \frac{M^2}{8} Tr (q^{-1} \nabla_\mu q )(q^{-1} \nabla^\mu q )  
\label{eff}
\end{eqnarray}
where spacetime indices, $\mu \nu$ are raised and lowered by the Minkowski  metric, $\eta^{\mu \nu}$. 

This effective action fails to be perturbatively renormalizable and thus it requires an ultraviolet completion.
One approach to this studied in \cite{ren} is to simply include all the dimension $4$ terms consistent with the symmetries.
This gives us,
\begin{eqnarray}
{\cal L}  &=&  -\frac{1}{4g^2_{1}} Tr   [q^{-1} F_{\mu \nu}^\dagger q F^{\mu \nu} ]
-\frac{1}{8g^{ 2}_{2}} Tr  [ F_{\mu \nu}  F^{\mu \nu} + F_{\mu \nu}^\dagger  F^{\dagger \mu \nu}  ]
\nonumber \\
&+& \frac{M^2}{8} Tr (q^{-1} \nabla_\mu q )(q^{-1} \nabla^\mu q )  -\frac{1}{8h} Tr (\nabla^2 q^{-1}  \nabla^2  q )
\nonumber \\
&-&\frac{\lambda_1}{32}[Tr (q^{-1} \nabla_\mu q )(q^{-1} \nabla^\mu q ) ]^2
- \frac{\lambda_2}{32}[Tr (q^{-1} \nabla_\mu q )(q^{-1} \nabla_\nu q ) ]^2
\label{ren}
\end{eqnarray}
 Note that the term proportional to $h^{-1}$ has higher derivatives, making the situation initially appear to be analogous to general relativity.   
But the situation is in fact different from general relativity, because the higher derivatives can be eliminated by an appropriate choice of gauge.  We will see this shortly.  

We note that the actions (\ref{eff},\ref{ren})  have a  symmetry, called $E$ parity:
\f
\omega_\mu \rightarrow -\omega_\mu^\dagger, \ \ \ \ q \rightarrow q^{-1}
\ff
This plays a crucial role in the following.

The internal  metric $q$ breaks the $SL(3,C)$ gauge symmetry into the $SU(3)$ symmetry that preserves $q$ and the 
transformations in $SL(3,C)/SU(3)$ that change $q$. 
The connection can be split into the metric compatible part, which is  antihermitian with respect to $q$, and so generates  $SU(3)$, called $A_\mu$, and the hermitian part with respect to $q$, $B_\mu$, which has components in 
$SL(3,C)/(SU(3))$.
\f
\omega_\mu = \imath A_\mu + B_\mu
\ff
where
\begin{eqnarray}
\imath A_\mu &=& \frac{1}{2} \left ( \omega_\mu - q^{-1} \omega_\mu^\dagger q
\right )
\\
B_\mu &=& \frac{1}{2} \left ( \omega_\mu + q^{-1} \omega_\mu^\dagger q
\right )
\end{eqnarray}
The action of $E$ parity is then,
\f
A_\mu \rightarrow q^{-1} A_\mu q , \ \ \ \ B_\mu \rightarrow - q^{-1} B_\mu q
\ff

Because $B_\mu$ is the matrix compatible part of the connection we have
\f
q^{-1} \nabla_\mu q = q^{-1} \partial_\mu q - 2 B_\mu
\ff
We write $F_{\mu \nu}$ in terms of these fields
\f
F_{\mu\nu}= \imath G_{\mu \nu} + W_{\mu \nu}
\ff
where
\f
G_{\mu \nu} = f_{\mu \nu} (A) + [B_\mu , B_\nu ], \ \ \ W_{\mu \nu}= {\cal D}_\mu B_\nu -  {\cal D}_\nu B_\mu
\ff
We can compute to find
\f
q^{-1} F_{\mu \nu}^\dagger q = -\imath G_{\mu \nu} + W_{\mu \nu} +[q^{-1} \nabla_\mu q , B_\nu -\imath A_\nu ]
-[q^{-1} \nabla_\nu q , B_\nu -\imath A_\mu ]
\ff
We can now rewrite the renormalizable lagrangian
\begin{eqnarray}
{\cal L}  &=&  -\frac{1}{4g^2} Tr   [G_{\mu \nu}G^{\mu \nu} ] - \frac{1}{4f^2} Tr   [W_{\mu \nu}W^{\mu \nu} ]
\nonumber \\
&+& \frac{M^2}{8} Tr (q^{-1} \partial_\mu q -2B_\mu )(q^{-1} \partial^\mu q -2 B^\mu)  -\frac{1}{8h} Tr (\nabla^2 q^{-1}  \nabla^2  q )
\nonumber \\
&-&\frac{\lambda_1}{32}[Tr (q^{-1} \partial_\mu q  -2B_\mu )(q^{-1} \partial^\mu q -2B^\mu ) ]^2
- \frac{\lambda_2}{32}[Tr (q^{-1} \partial_\mu q  -2B_\mu )(q^{-1} \partial_\mu q  -2B_\nu ) ]^2
\nonumber \\
& -& \frac{1}{4g^2_{1}} Tr   [ (\imath G^{\mu \nu} + W^{\mu \nu} ) (q^{-1} \partial_\mu q  -2B_\mu, B_\nu -\imath A_\nu ]
-[q^{-1} \partial_\mu q  -2B_\mu , B_\nu -\imath A_\mu ] )]
\nonumber \\ 
&-& \frac{1}{4g^{2}_{2}} Tr  [ -\imath G^{\mu \nu} + W^{\mu \nu}   ]
[[q^{-1} \partial_\mu q  -2B_\mu , B_\nu -\imath A_\nu ]
-[q^{-1} \partial_\mu q  -2B_\nu , B_\nu -\imath A_\mu ]  ]
\nonumber \\ 
&-& \frac{1}{8g^{2}_{2}} Tr  [[q^{-1} \partial_\mu q  -2B_\mu, B^\nu -\imath A^\nu ]
-[q^{-1} \partial_\nu q  -2B_\mu , B_\nu -\imath A^\mu ]  ] 
\nonumber \\ 
&&\times
[[q^{-1} \partial_\mu q  -2B_\mu , B_\nu -\imath A_\nu ]
-[q^{-1} \partial_\nu q  -2B_\nu , B_\nu -\imath A_\mu ]  ]
\label{ren3}
\end{eqnarray}

where we have redefined the coupling constants
\f
\frac{1}{g^2}= \frac{1}{g^2_{1}}+ \frac{1}{g^{2}_{2}} , \ \ \ \  \frac{1}{f^2}=  \frac{1}{g^2_{1}}-  \frac{1}{g^{2}_{2}} .
\ff
$g^2 $ is the $QCD$ coupling constant, while $f$ is a new coupling constant.

Both $g^2$ and $f^2$ have to be positive definite.  A calculation \cite{ren} shows that both are asymptotically free in the absence of fermions.

When we come to the quarks we note that the standard coupling,
\f
{\cal L}^\Psi = \bar{\Psi} \gamma^\mu {\cal \nabla}_\mu \Psi + h.c. 
\ff
involves only the $A_\mu$.  Indeed it appears that there are no couplings of the $E$-bosons to chiral fermions, consistent with $E$-parity, at least at low order.

If we want the $E$'s to couple to fermions preserving $E$-parity,  we have to introduce new, vector-like or non-chiral quarks.  Assuming we do so, the leading direct coupling between the $E$'s and the new, vector quarks, will arise quadratically through dimension $5$ terms, which preserve $E$-parity. the simplest of which is,
\f
{\cal L}^\Psi_{(5)} =   \frac{c_0}{M} \bar{\Psi}  (q^{-1} \nabla_\mu q )  (q^{-1} \nabla^\mu q ) \Psi  
 \label{dim5}
\ff
This introduces a parameters $c_0$.  which is a function of the original coupling constants,
$g$ and $f$ as they arise from one loop diagrams.  A simple calculation\cite{inprogress} shows that 
\f
c_0 \approx \frac{1}{\pi^2} g^2 (g^2 +f^2 ) \frac{m_q}{M}
\label{cf}
\ff
where $m_q$ is the vector quark mass.
As $f$ is a new parameter, it can be traded for $c_0$.  

We now can fix the $SL(3,C)/(SU(3))$ part of the gauge invariance, by setting,
\f
\partial_\mu q  = 0.
\ff
This is a consistent gauge choice because in any gauge the $q$ field equation is equal to the covariant divergence 
of the $B_\mu$ field equations\cite{JJL,ren}.  
This leaves a global $SL(3,C)$ symmetry but breaks the local symmetry down to $SU(3)$.  We note that
\f
q^{-1} \nabla_\mu q|_{\partial_\nu q =0} = -2 B_\mu
\ff

$E$-parity now becomes
\f
B_\mu \rightarrow -B_\mu, \ \ \ \ A_\mu \rightarrow A_\mu
\ff

The $E$-parity invariant action now greatly simplifies to
\begin{eqnarray}
{\cal L}  &=&  -\frac{1}{4g^2} Tr    [f_{\mu \nu} (A) + [B_\mu , B_\nu ]]^2 - \frac{1}{4f^2} Tr   [W_{\mu \nu}W^{\mu \nu} ]
\nonumber \\
&+& \frac{M^2}{2} Tr (B_\mu  B^\mu)  +\frac{1}{8h} Tr ({\cal D}_\mu B^\mu )^2
\nonumber \\
&-&\frac{\lambda_1^\prime }{4}[Tr (B_\mu B^\mu ) ]^2
- \frac{\lambda_2}{4}[Tr (B_\mu  B_\nu  ) ]^2
\label{ren4}
\end{eqnarray}
where 
\f
\lambda_1^\prime =\lambda_1 - \frac{8}{h}
\ff

\section{The physics}

The quantization of this theory, with and without the $h^{-1}$ term, is discussed in detail in \cite{JJL,ren,others-q}.  

The $1/h$ term gives a kinetic energy to $B_0$.  That gives us four degrees of freedom per gauge generator of $SL(3,C)/SU(3)$.
In the 
$\partial_\mu q=0$ gauge the $B$- propagator has the form\cite{ren}
\f
D_{\mu\nu} (p^2 ) =  \imath \frac{\eta_{\mu \nu} -\frac{p_\mu p_\nu }{p^2}  }{p^2 + M^2 + \imath \epsilon} +
\frac{\imath  \frac{p_\mu p_\nu }{p^2}  }{\frac{p^2}{h} + M^2 + \imath \epsilon}
\ff
We choose  $h>0$ in which  case there is neither a longitudinal ghost nor a  tachyon.  We can choose to take  $h$ very large so that the longitudinal pole occurs at much higher than $LHC$ energies.  

The theory given by eq. (\ref{ren3}) is renormalizable and indeed asymptotically free in both $g^2$ and $f^2$, 
but there are issues of stability of the perturbative ground state, as discussed in \cite{ren}.  
These issues are slightly analogous to the issues that arise in the Stelle renormalizable extension of $GR$ but not precisely the same as there are no higher derivatives in this gauge.
These issues may be resolvable by asymptotic safety or the Lee-Wick mechanism.  But as these are presently unresolved, here 
we are going to take the limit $h \rightarrow \infty $ and avoid these issues.  We will treat the theory given by 
eq (\ref{eff}) as an effective field theory with an unknown ultraviolet completion. 

When $ p^2 << -h M^2$ the propagator becomes the one of the effective theory,
\f
D_{\mu\nu} (p^2 ) =  \imath \frac{\eta_{\mu \nu} -\frac{p_\mu p_\nu }{p^2}  }{p^2 - M^2 + \imath \epsilon} +
\frac{\imath  \frac{p_\mu p_\nu }{p^2}  }{  M^2 }. 
\ff


The theory we will consider is the effective action including the dimension five couplings of $B$'s to quarks.

\begin{eqnarray}
{\cal L}  &=&  -\frac{1}{4g^2} Tr    [f_{\mu \nu} (A) + [B_\mu , B_\nu ]]^2 - \frac{1}{4f^2} Tr   [W_{\mu \nu}W^{\mu \nu} ]
\nonumber \\
&+& \frac{M^2}{2} Tr (B_\mu  B^\mu) 
+
 \bar{\Psi} \gamma^\mu {\cal D}_\mu \Psi + h.c. + m_q \bar{\Psi}{\Psi}
+ \frac{c_0}{M}  \bar{\Psi} B_\mu B^\mu \Psi  
\label{ren4}
\end{eqnarray}

Below the mass scale $M$ this is $QCD$.  We have an octet of coloured massive vector bosons, $B_\mu$, which
come from the $SL(3,C)/SU(3)$ gauge fields.  Thus we have a mechanism analogous to the Higgs mechanism in which the low energy symmetry has spontaneously broken from $SL(3,C)$ to $SU(3)$.  We then have an elegant extension of $QCD$
which adds an octet of heavy gluons in a way that is controlled by one dimensionless constant and one mass.


The gluons couple to the $B_\mu$'s by several terms.  $E$-parity enforces that all of these are quadratic in $B_\mu$.

There is the cubic coupling 
\f
(g+\frac{f^2}{g}) Tr ((  \partial_\mu B_\nu -  \partial_\nu B_\mu   ) [A^\mu , B^\nu ] ) 
\ff
and the quartic coupling by which a pair of gluons converts to a pair of $B_\mu $'s.
\f  
g^2 (1 + \frac{f^2}{g^2} ) Tr [A^\mu , B^\nu ]^2   
\ff
There is finally a term by which a pair of $E$'s can convert to a pair of quarks
\f
\frac{c_0}{M} \Psi^\dagger  B_\mu B^\mu \Psi  
\ff

\section{Could this explain the new particle at the LHC?}

The $E$ boson is a massive coloured octet, thus it cannot appear as a final state but must be confined into coloured singlets.
There are three kinds of bound states we have to  consider.

\begin{figure*}[t]
\begin{center}
\includegraphics[scale=0.6]{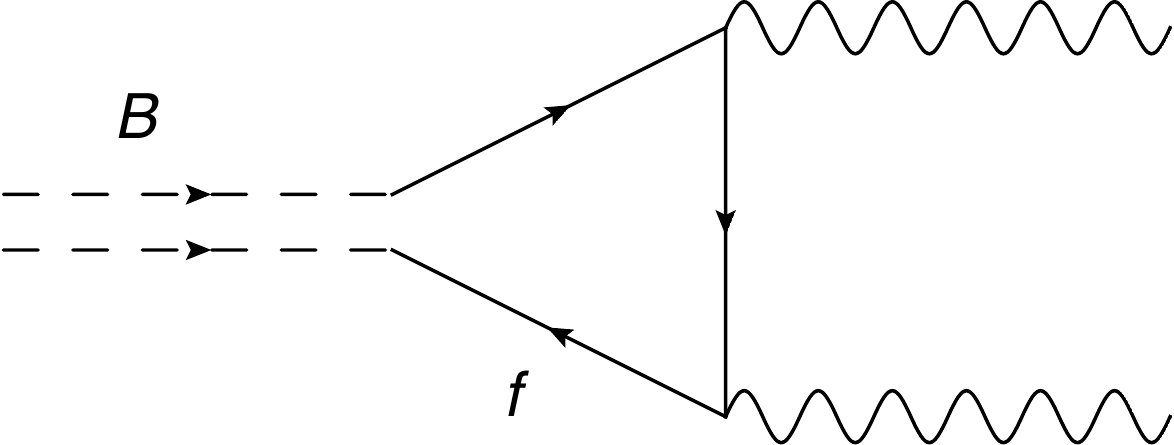}
\caption{Production of two photons from a bound scalar state through a quark loop}
\end{center}
\end{figure*}

\begin{itemize}

\item{} Bound states of two $E$'s.  They can  have spin zero or spin two.  The spin $0$ particles have even parity, we will call them $S$.  They have masses around $(2 -\alpha_S^2 ) M$ and are neutral.  These have zero  $E$-parity and can decay directly back to pairs of gluons.  These can also decay to pairs of photons via the dimension 5 coupling shown in eq (\ref{dim5}), followed by a vector quark loop as shown in Figure 1.  They are candidates for the diphoton resonance.  

The production and decay of these bound states should be similar to quarkonium, the bound states of 
heavy quarks\cite{octet,gluinos}.  If this is the correct interpretation, we expect a spectrum of excited states with spacings of order 
$\Delta M \approx \alpha_S^2 M \approx 4 Gev$.

The decay of these mesons into photons goes through a loop of vector quarks through the dimension $5$ couplings in the lagrangian (\ref{dim5},\ref{ren4}), as shown in Figure 1.  This is controlled at leading order by the coupling constant, $c_0$
(which is determined by the new constant $f$, through (\ref{cf}).).   

The lowest order decay rate of the $E$'s to two photons via a quark loop will be roughly\cite{octet,project,inprogress}:
\be 
\Gamma(S \rightarrow \gamma\gamma)  \sim \frac{3}{16\pi}c^{2}_0 \alpha^3_S \alpha^{2}  \frac{m_q^2}{M}
\label{gammadecay}
 \ee
 where $\alpha_S \approx .1$ at these scales.  The factor $\alpha_S^3$ comes from the wave function at the origin.
 \f
\lambda_M^3  |\Psi (0) |^2 = \frac{\lambda_M^3}{a_{Bohr}^3} = \alpha_S^3
\label{Gammagg}
 \ff
The question is whether the vector quark mass $m_q$ can be chosen so that this process is not too much dominated by the decay back through a pair of gluons, which is roughly,
\be 
\Gamma(S \rightarrow \ g g )  \sim (g^2 + f^2 )^2 \alpha_S^3 M .
\label{gluondecay}
 \ee
 
To see whether this  can be achieved, we note the following.  First, $ATLAS$
has just put a lower bound of $4.4 \ TeV$ on the mass of any new vector quark\cite{limitsonquarks}.
Thus
\f
\frac{m_q^2}{M^2 } > 137
\label{mq}
\ff

However, the measured (if the state is real) partial cross section of two gluons to $S$ and back to two photons, is\cite{ATLAS,CMS}
 \f
 \sigma (gg \rightarrow S )  B_{S \rightarrow \gamma \gamma }\approx 5 \ fb
 \label{measured}
 \ff
 At the same time there are recent limits from $CMS$ on the partial cross section to two gluons\cite{scouting}, 
  \f
 \sigma (gg \rightarrow S )  B_{S \rightarrow  g g } < 1000 \ fb
 \ff
 The decay through gluons is expected to dominate, and indeed we see from (\ref{measured}) and (\ref{gfusion}) that the ratio 
 \f
 R = \frac{\Gamma(S \rightarrow \gamma\gamma)  }{\Gamma(S \rightarrow \ g g ) } > \frac{1}{200}
 \ff
 This agrees with an estimate for the rough value of the cross section for gluon fusion to $S$ is
   \f
 \sigma (gg \rightarrow S )  = (g^2 + f^2 )^2 \alpha_S^3 M^2  > 10 fb  (1 +\frac{f^2}{g^2} )^2
 \label{gfusion}
 \ff
 where $(g^2 + f^2 )^2 $ comes from the vertices and $\alpha_S^3$ comes from the wave function at the origin.
 
 But we have, from (\ref{gammadecay}) and (\ref{gluondecay})
 \f
 R = \frac{\Gamma(S \rightarrow \gamma\gamma)  }{\Gamma(S \rightarrow \ g g ) } \approx 3 \cdot c_0^2 
 \frac{m_q^2}{M^2} 
 \frac{\alpha^2 }{(g^2 + f^2 )^2}
 \ff
this implies, using (\ref{cf}) that
 \f
R \approx \frac{3}{\pi^4} \frac{m^4_q}{M^4} \alpha^2 \alpha^2_s > \frac{1}{200}
 \ff
 This implies very roughly that
 \f
 \frac{m_q}{M} > 20
 \label{mqbound}
 \ff
 which is consistent with the observed bound (\ref{mq}) that $ \frac{m_q}{M} > 12 $.

 
 Finally, we see from (\ref{Gammagg}) that the width is roughly $\Delta \approx .5 MeV $ which is indeed narrow.


\item{}Single $B_\mu$ states with $E$-parity equal to one and no quarks.  These  can be thought of as a bound state of a 
$E$ with a gluon.  They can  have spin zero or spin two.  They have masses around $M$ and are neutral.

By E-parity, these are stable and are hence dark matter candidates.  They are not candidates for the diphoton resonance.

As strongly interacting dark matter candidates they face strict limits, and cannot be a dominant component of dark matter. Their density will be lowered by the fact that they annihilate in pairs.  It should be possible to estimate the relic density and see if it is under the existing limits.

\item{}Single $B_\mu$ states with $E$-parity equal to one and a  quark antiquark pair.  These  can be thought of as a bound state of a 
$E$ with a quark and an antiquark.  They also have masses around $M$ and electroweak charges.

These are like very heavy mesons and decay to neutrinos and muons and the first kind.


\end{itemize}

\section{Conclusions}

In this paper we have put forward a tentative hypothesis for the new state at $750 \ GeV$ possibly seen at the $LHC$.
This is that it is a colour singlet bound state of pair of heavy gluons.  We impose a conservation law, $E$-parity, which requires
that the heavy gluons can only be created or annihilated in pairs.  This forces their coupling to quarks to go through
a dimension $5$ operator.  Further, only a new generation of vector quarks can couple directly to the $E$'s, preserving $E$-parity.    We argued that the resulting physics is at least roughly consistent with what is known of the 
new resonance.  We see this from the fact that we can roughly match the observed cross sections and limits with a value 
for the dimensionless parameter of the dimension $5$ coupling of the bound state to vector quarks of order unity.


We then discussed one very natural extension of the standard model which realizes these hypotheses.
This is based on an extenson
of the color  $SU(3)$ gauge invariance to $SL(3,C)$, consistent with maintaining the principle that the energy is bounded from below.   This kind of extension was proposed in the early 1980's as a route to unifying gauge theory with 
gravity\cite{Cahill,Dell,Zee,Julia}, as they involve a dynamical internal metric.  This is an example of a possible consequences of a unification of gravity with particle physics.  


\section*{Acknowledgements}

We are grateful to Miriam Diamond, Michael Luke, Marcus Spradlin,   Matthew Strassler and Devin Walker for very helpful conversations and to Jacob Barnett, Kevin Cahill and Michael Peskin, for comments on drafts of this paper.   This research was supported in part by Perimeter Institute for Theoretical Physics. Research at Perimeter Institute is supported by the Government of Canada through Industry Canada and by the Province of Ontario through the Ministry of Research and Innovation. This research was also partly supported by grants from NSERC, FQXi and the John Templeton Foundation.


\begin{thebibliography}{99}

\bibitem{ATLAS}G. Aad et al. ATLAS Collaboration, Note ATLAS-CONF-2015-081, CERN, Geneva (2015), URL http://cds.cern.ch/record/2114853.

\bibitem{CMS}V. Khachatryan et al. 
 CMS Collaboration, Note CMS-PAS-EXO-15-004, CERN, Geneva (2015), URL https://cds.cern.ch/record/2114808.

\bibitem{Cahill}K. Cahill, Phys. Rev. D 18, 2930 (1978); 20; 2636 (1979); J.
Math. Phys. 21, 2676 (1980); Phys. Rev. D 26, 1916 (1982).

\bibitem{Dell}J. C. Dell, University of Maryland report, 1979 (unpublished); Ph.D dissertation 1980.

\bibitem{Zee}J. E. Kim and A. Zee, Phys. Rev. D. 21, 1939 (1980); 

\bibitem{Julia}B. Julia
and F. Luciani, Phys.Lett.908,270(1980)


\bibitem{DellSmolin} John Dell and Lee Smolin, Composite Higgs bosons from an extended gauge symmetry, Enrico Fermi Institute preprint, December 1983, revised, December 1984.

\bibitem{JJL}John Dell, Jorge L. deLyra, Lee Smolin, {\it Quantization of a Gauge Theory With Independent Metric and Connection Fields}, Phys.Rev. D34 (1986) 3012. 

\bibitem{ren}Lee Smolin, {\it  A new class of renormalizable and asymptotically free gauge theories},  Phys. Rev. D 30, 2159 (1984).

\bibitem{others-q}W. Ishizuka, Y. Kikuchi,  {\it 
Beta Functions Of The Four Point Coupling Constants In A New Class Of Gauge Theory}, 
Phys.Rev. D33 (1986) 1837-1838; Shinichi Deguchi,  {\it Quantization Of Gauge Theory With Scalar Field Dependent Metric}, Prog.Theor.Phys. 82 (1989) 433-448. 

\bibitem{Plebanski} 
  J.~F.~Plebanski,
  {\it On the separation of Einsteinian substructures,}
  J.\ Math.\ Phys.\  {\bf 18}, 2511 (1977).


\bibitem{Ashtekar} 
  A.~Ashtekar,
  {\it New Variables for Classical and Quantum Gravity,}
  Phys.\ Rev.\ Lett.\  {\bf 57}, 2244 (1986).

\bibitem{IK}W. Ishizuka, Y. Kikuchi , {\it Beta Functions Of The Four Point Coupling Constants In A New Class Of Gauge Theory},  Phys.Rev. D33 (1986) 1837-1838; Shinichi Deguchi, S. Deguchi, {\i\t Quantization Of Gauge Theory With Scalar Field Dependent Metric},  Prog.Theor.Phys. 82 (1989) 433-448.  

 \bibitem{Alexander:2007mt} 
  S.~H.~S.~Alexander,
  {\it Isogravity: Toward an Electroweak and Gravitational Unification,}
  arXiv:0706.4481 [hep-th];
  F.~Nesti and R.~Percacci,
  {\it Graviweak Unification,}
  J.\ Phys.\ A A {\bf 41} (2008) 075405,
  arXiv:0706.3307 [hep-th];  S.~Alexander, A.~Marciano and R.~A.~Tacchi,
  {\it Towards a Loop Quantum Gravity and Yang-Mills Unification,}
  Phys.\ Lett.\ B {\bf 716} (2012) 330
  arXiv:1105.3480 [gr-qc].
  
\bibitem{graviweak} Stephon Alexander, Antonino Marciano, Lee Smolin, {\it Gravitational origin of the weak interaction's chirality,}  arXiv:1212.5246 [hep-th], Phys. Rev. D 89, 065017 (2014).  

\bibitem{AS}Weinberg, STeVen (1979). Ultraviolet divergences in quantum theories of gravitation. In "General Relativity: An Einstein centenary survey", ed. S. W. Hawking and W. Israel. Cambridge University Press. pp. 790?831; Parisi, Giorgio (1976). "On Non-Renormalizable Interactions". New Developments in Quantum Field Theory and Statistical Mechanics Cargèse. doi:10.1007/978-1-4615-8918-1 12; M. Reuter and F. Saueressig, 
{\it Quantum Einstein Gravity,}
New J.\ Phys.\ {\bf 14}, 055022 (2012)  
[arXiv:1202.2274 [hep-th]].

\bibitem{LW}T. D. Lee and G. C. Wick, Phys. Rev. D 2, 1033 (1970) , Nucl. Phys. B 9, 209 (1969) and
Phys. Rev. D 3, 1046 (1971).  See also  B. Grinstein, D. O?Connell and M. B. Wise, arXiv:0704.1845 [hep-ph].
  
 \bibitem{diagonal}Jia Liu, Xiao-Ping Wang, and Wei Xue, {\it  LHC diphoton excess from colorful resonances },   ArXiv:1512.07885.
 
 \bibitem{MattS1}Prateek Agrawal, JiJi Fan, Ben Heidenreich, Matthew Reece, Matthew Strassler, {\it Experimental Considerations Motivated by the Diphoton Excess at the LHC },
 arXiv:1512.05775.
 
 \bibitem{random}Wolfgang Altmannshofer, Jamison Galloway, Stefania Gori, Alexander L. Kagan, Adam Martin, Jure Zupan, 
 {\it On the 750 GeV di-photon excess},  arXiv:1512.07616; 
S.~Kanemura, N.~Machida, S.~Odori and T.~Shindou, {\it Diphoton excess at 750 GeV in an extended scalar sector,'}
arXiv:1512.09053 [hep-ph]; 
  Q.~H.~Cao, Y.~Liu, K.~P.~Xie, B.~Yan and D.~M.~Zhang,
  {\it A Boost Test of Anomalous Diphoton Resonance at the LHC,'}
  arXiv:1512.05542 [hep-ph]; 
  Q.~H.~Cao, Y.~Liu, K.~P.~Xie, B.~Yan and D.~M.~Zhang,
{\it The Diphoton Excess, Low Energy Theorem and the 331 Model,'}
  arXiv:1512.08441 [hep-ph];  Ilja Dorsner, Svjetlana Fajfer, Nejc Kosnik,   {\it Is symmetry breaking of SU(5) theory responsible for the diphoton excess?   },  arXiv:1601.03267. 



\bibitem{MattS2}Yevgeny Kats and Matthew J. Strassler, {\it Probing Colored Particles with Photons, Leptons, and Jets}, 
  arXiv:1204.1119.
 
 \bibitem{octet}Chul Kim, Thomas Mehen,  {\it Color Octet Scalar Bound States at the LHC},  	10.1103/PhysRevD.79.03511, arXiv:0812.0307,  http://arxiv.org/abs/0812.0307 

\bibitem{gluinos}E.Chikovani, V.Kartvelishvili, R.Shanidze, G.Shaw, {\it Bound States of Two Gluinos at the TeVatron and LHC}, Phys.Rev.D53:6653-6657,1996, http://arxiv.org/abs/hep-ph/9602249.

\bibitem{scouting}CMS Collaboration, {\it Search for resonances decaying to dijet final states at 8 TeV with scouting data,}
CMS-PAS-EXO-14-005 (2014) . https://cds.cern.ch/record/2063491

\bibitem{limitsonquarks}ATLAS collaboration, {\it Search for new phenomena with photon+jet events in
proton?proton collisions at ?s = 13 TeV with the ATLAS detector },  CERN-PH-EP-2015-320,  http://arxiv.org/pdf/1512.05910v1.pdf

\bibitem{project}Peskin, Michael E., and Daniel V. Schroeder. An introduction to quantum field theory. Westview, 1995; Christopher Schwan, http://www.students.uni-mainz.de/cschwan/higgs-final-project.pdf

\bibitem{inprogress}S. Alexander and L. Smolin in progress.

\end{thebibliography}
\end{document}